\documentclass[10pt,journal]{IEEEtran}
\ifCLASSINFOpdf
\else
\fi
\usepackage{optidef}
\patchcmd{\bibsetup}{\interlinepenalty=5000}{\interlinepenalty=0}{}{}
\allowbreak
\usepackage{titlesec}
\usepackage[skip=0pt]{caption} 
\allowdisplaybreaks
\usepackage{braket}
\usepackage{lettrine}

\usepackage{etoolbox}
\appto{\bibsetup}{\raggedbottom}

\usepackage[labelformat=simple]{subcaption}  

\usepackage{algorithm}
\usepackage{placeins}
\usepackage{comment}

\usepackage{algorithm}
\usepackage{algpseudocode}
\usepackage[english]{babel}

\usepackage{mathtools}
\usepackage{soul}
\usepackage{pdfpages}

\usepackage[skip=3pt]{caption}

\usepackage{etoolbox}
\AfterEndEnvironment{algorithm}{\vspace{-3pt}}

\AtBeginDocument{%
  \setlength\abovedisplayskip{6pt}
  \setlength\belowdisplayskip{6pt}
  \setlength\abovedisplayshortskip{3pt}
  \setlength\belowdisplayshortskip{3pt}
}

\usepackage{titlesec}
\titlespacing*{\section}{0pt}{6pt plus 2pt minus 1pt}{3pt plus 1pt minus 1pt}
\titlespacing*{\subsection}{0pt}{5pt plus 1pt}{2pt plus 1pt}
\titlespacing*{\subsubsection}{0pt}{4pt plus 1pt}{2pt plus 1pt}

\makeatletter
\def\thm@space@setup{%
  \thm@preskip=3pt
  \thm@postskip=3pt
}
\makeatother

\usepackage{tabularx,booktabs}
\usepackage {xcolor}
\usepackage[utf8]{inputenc}
\usepackage{amssymb}
\usepackage{etoolbox}
\AtBeginDocument{%
  \setlength\abovedisplayskip{6pt}
  \setlength\belowdisplayskip{6pt}
  \setlength\abovedisplayshortskip{3pt}
  \setlength\belowdisplayshortskip{3pt}
}
\usepackage{color}
\usepackage{graphicx,amssymb,amsfonts,booktabs,multirow,array,comment}

\usepackage[all,graph]{xy}

\makeatletter
\def\thm@space@setup{%
  \thm@preskip=0pt
  \thm@postskip=0pt
}
\makeatother

\usepackage{mathtools}
\begin{document}
\bstctlcite{IEEEexample:BSTcontrol}
\title{Quantum Graph Neural Networks for Double-Sided Reconfigurable Intelligent Surface Optimization}

\author{Noha~Hassan,~\IEEEmembership{SMIEEE,}
        Xavier~Fernando,~\IEEEmembership{SMIEEE,}
        and~Halim~Yanikomeroglu,~\IEEEmembership{FIEEE}

\thanks{This work has been submitted to the IEEE Wireless Communications Letters Journal for possible publication. Copyright may be transferred without notice, after which this version may no longer be accessible.}
\thanks{N. Hassan and H. Yanikomeroglu are with the Department of Systems and Computer Engineering,
Carleton University, Ottawa, ON, Canada
(e-mail:  noha.hassan@torontomu.ca; halim@sce.carleton.ca).}
\thanks{X. Fernando is with the Department of Electrical, Computer, and Biomedical Engineering,
Toronto Metropolitan University, Toronto, ON, Canada
(e-mail: fernando@torontomu.ca).}}

\maketitle

\markboth{IEEE Wireless Communications Letters}%
{Hassan \MakeLowercase{\textit{et al.}}: }

\pagestyle{headings}
\vspace{-20pt}
\begin{abstract}
As a key enabler for sixth-generation (6G) wireless communications, reconfigurable intelligent surfaces (RISs) provide the flexibility to control signal strength. Nevertheless, optimizing hundreds of elements is computationally expensive. To overcome this challenge, we present a quantum framework (QGCN) to jointly optimize the physical and electromagnetic response of a double-sided RIS design that incorporates discrete phase shifts and inter-element coupling. The core contribution is the adaptive activation or deactivation of elements, allowing a virtual spacing mechanism using PIN diode switches. We then solve a multi-objective problem that maximizes the minimum user data rate subject to constraints on aperture length and mutual coupling between active elements. 
Experimental results on IBM Quantum's 127-qubit \textit{ibm\_kyiv} superconducting processor demonstrate that the proposed QGCN algorithm reduces both per-iteration computational complexity and memory requirements compared to existing approaches. Also, the QGCN outperforms classical graph neural networks (GNN) on an equivalent graph topology by an additional +0.38 bps/Hz. This advantage is increasing with increasing array sizes.    
\end{abstract}
\IEEEpeerreviewmaketitle
\vspace{-6pt}
\section{Introduction}
\lettrine{R}{econfigurable} intelligent surfaces (RISs) are a leading enabler of sixth-generation (6G) systems, supporting hyper-reliable low-latency communication (HRLLC), massive communication, and immersive communication \cite{itur_m2516_2022}. RISs enable substantial gains in capacity, energy efficiency, and service quality. An  RIS has electronically tunable reflectors that dynamically manage electromagnetic propagation through control of reflection coefficients and phase shifts \cite{iqbal2025comprehensive}.
To further boost the flexibility of the RIS design, we consider a double-sided structure with environment-based activation. The two planar sides of the RIS are independently activated or deactivated depending on propagation geometry, user (UE) distribution, and network traffic. This structure enables the RIS panel to dynamically switch its active reflecting surface to the dominant direction of communication, which could be forward, backward, or bidirectional. The design also reduces interference and power consumption while maintaining channel controllability.
Most existing works treat element spacing as constant and only optimize phase shifts \cite{xiao2025fluid}. This relaxation overlooks the key performance impacts of element positioning on array-factor design, path-loss scaling, and inter-service interference. Also, the coherent combining of large-scale elements makes the received power a function of the complex superposition of element contributions, leading to highly non-convex objective spaces. The two-way propagation relates spatial placement to receiver signal quality in a nonlinear manner. To address these challenges, we propose a multi-objective framework that jointly optimizes RIS element spacing and phase shifts.
In practical wireless communication systems, other sources of performance degradation, such as phase noise, quantization effects, and transceiver impairments, are also present, as discussed in~\cite{li2025stacked}. These effects are also accounted for in the proposed model through discrete phase shifts, mutual coupling effects, and binary PIN-diode switching. Although an explicit hardware impairment model is beyond the scope of this work, the proposed formulation captures the dominant effects on RIS performance.
Our contributions are (i) an end-to-end system model of spacing-dependent channels with mutual coupling and multi-service operation using varactor-based tunable elements; (ii) a multi-objective analytical formulation for phase-spacing joint optimization; (iii) the QGCN algorithm for large-scale RIS networks; (iv) a double-sided architecture for selective element activation; (v) element spacing variation is implemented using switch-based virtual spacing using PIN diode RF switches on a dense grid; and (vi) validation on noisy intermediate-scale quantum (NISQ) IBM quantum hardware. 

The remainder of this paper is organized as follows: Section II reviews related work. Section III is the system model. Section IV presents the algorithm description. Section V provides the results. Finally, Section VI concludes the paper.
\vspace{-0.5em}
\section{Literature Review}
Classical optimization algorithms have significant limitations of slow convergence and challenges to handle non-convex objective functions and constraints, and they relax phase discrete variables to continuous variables in large-scale RIS element optimization.
Authors in \cite{li2023low} jointly optimize the number of reflecting elements at a reconfigurable intelligent surface and their phase shifts to maximize the system capacity while accounting for channel estimation overhead and phase quantization errors. 
However, they focus on single-frequency scenarios, do not address mutual coupling effects between elements, and assume static channel state information (CSI) rather than dynamic user mobility.
Gradient-based topology optimization techniques for RIS are applied in \cite{karamanos2024topology}. It addresses the binary decisions of element locations in an orderly fashion, providing an electromagnetic (EM)-guided inverse design strategy. Computational complexity can be prohibitive, and scalability to massive arrays remains an open problem.
The work in \cite{tseng2024irregular} considers mutual coupling effects and illustrates that optimally chosen subsets of elements can outperform full whole-form arrays. The paper provides some insight into effective element selection and energy-saving RIS design. 
More recently, double-sided RIS architectures have been proposed to increase the coverage and system flexibility by working in both reflection and transmission modes. Novel designs focus on signal modeling, mode and phase optimizations under fixed element geometry. The study in~\cite{gao2025dual} demonstrates the potential of double-sided RISs to improve spectral efficiency, but its design is based on a fixed inter-element spacing, centralized optimization, and continuous phase models under static configurations, while paying limited attention to online adaptability.
Deep neural networks (DNNs) and graph neural networks (GNNs) have been used to learn phase shift mappings from CSI with reduced computational cost~\cite{shen2023deep,chen2023graph}.
Quantum optimization has also been proposed for wireless systems, but joint phase–spacing RIS optimization with hardware validation remains largely unexplored~\cite{chehimi2025reconfigurable}.
Three important gaps exist in the literature. First, the majority of previous works optimize phase shifts or surface topology alone, rather than jointly, particularly in double-sided RISs where front- and back-side interactions are closely related to each other. Second, traditional optimization techniques and gradient-based methods tend to be computationally infeasible due to the discrete, non-convex, and high-dimensional nature of joint spacing and phase optimization problems, especially with time-varying UE distributions. Most of the previous two-sided RIS works are based on static schemes and perfect CSI.

This paper fills the gaps in literature by using a quantum graph neural network (QGNN) that: (i) jointly optimizes space and phase by dual-qubit encoding without continuous relaxation, (ii) handles non-convex discrete-continuous spaces by structured quantum graph representations, (iii) validates on IBM Quantum hardware (not just simulation), and (iv) supports online adaptation for dynamic double-sided RIS deployment.
\vspace{-0.9em}
\section{System Model}
\subsection{Physical Model}
\vspace{-1mm}
The system model considers a wireless network consisting of $M$ access points (APs) to support $K$ UEs. Moreover, an adaptive double-sided RIS consisting of $N$ passive reflector elements with activation-controlled effective inter-element spacing is included in the network. The above setup enables front or back reflective surface activation based on UE distribution, propagation conditions, or traffic demand. 
Each RIS element has a phase shift $\phi_n \in [0,2\pi)$ with the incoming signal.
The cascaded channel reflects whether signals propagate via the front side, the back side, or both, depending on the relative positions of APs and UEs.
The elements are arranged on a dense fixed grid with uniform minimum spacing $d_{\min} = \lambda/2$, where $\lambda$ is the operating wavelength. The position of element $n$ is $p_{n} = (n-1) d_{\min}$.
Each element is equipped with a PIN diode switch that enables rapid activation ($a_n = 1$) or deactivation ($a_n = 0$). The system creates virtual spacing patterns by selectively activating a subset of elements without any physical movement of the substrate.
The activation pattern vector $\boldsymbol{a} = [a_1, a_2, \dots, a_{N}]^{\mathsf{T}}$ determines which elements contribute to the reflected signal, while deactivated elements ($a_n = 0$) effectively become transparent to the incident wave.
\subsection{Signal and Channel Model}

We adopt the Rician fading channel model, and the proposed framework is based on time-division duplexing (TDD) and channel reciprocity, where the uplink pilot transmission allows to estimate the cascaded AP--RIS--UE channel. 
The channel between AP $m$ and RIS element $n$ located at position $p_n$ is expressed as
\vspace{-0.7em}\begin{equation}
h_{m,n}(p_n) = \sqrt{\frac{\kappa}{\kappa+1}}\, h_{m,n}^{\text{LoS}}(p_n) + \sqrt{\frac{1}{\kappa+1}}\, h_{m,n}^{\text{NLoS}}(p_n),
\end{equation}
where $\kappa$ denotes the Rician factor.
The LoS depends on the element position and is expressed as  
$h_{m,n}^{\text{LoS}}(p_n) = \frac{\lambda}{4\pi r_{m,n}} e^{-j\frac{2\pi}{\lambda}r_{m,n}} e^{j\frac{2\pi}{\lambda}p_n\sin\theta_m}$,
where $r_{m,n}$ represents the distance between AP transmit antenna $m$ and RIS element $n$, and $\theta_m$ is the angle of departure. 
The non-line-of-sight (NLoS) component is characterized as a circularly symmetric complex Gaussian random variable, i.e., $h_{m,n}^{\text{NLoS}} \sim \mathcal{CN}(0,1)$.

Similarly, the RIS-to-UE channel is defined as  
\begin{equation}
h_{n,k}(p_n) = \sqrt{\frac{\kappa}{\kappa+1}}\, 
h_{n,k}^{\text{LoS}}(p_n) + \sqrt{\frac{1}{\kappa+1}}\, h_{n,k}^{\text{NLoS}}(p_n),
\end{equation}
where the LoS component is given by $h_{n,k}^{\text{LoS}}(p_n) = \frac{\lambda}{4\pi r_{n,k}} e^{-j\frac{2\pi}{\lambda}r_{n,k}} e^{-j\frac{2\pi}{\lambda}p_n\sin\theta_k}$, 
with $\theta_k$ denoting the angle of arrival at UE~$k$.
The overall cascaded channel from AP~$m$ to UE~$k$ through the adaptive RIS is expressed as 
$h_{m,k}(\boldsymbol{\Phi}, \boldsymbol{a})
= \sqrt{\beta_k}\,
\mathbf{h}_{m}^{\mathsf{H}}(\boldsymbol{a})
\mathbf{C}(\mathbf{d}(\boldsymbol{a}))
\mathbf{R}(\boldsymbol{\Phi}, \boldsymbol{a})
\mathbf{h}_{k}(\boldsymbol{a})$,
where $h_{m,k}(\boldsymbol{\Phi}, \boldsymbol{a}) \in \mathbb{C}$ is a scalar channel coefficient, $\beta_k$ represents the composite large-scale path loss (AP→RIS and RIS→UE), and $\mathbf{R}(\boldsymbol{\Phi}, \boldsymbol{a}) = \mathrm{diag}\{a_{1}e^{j\phi_1}, \ldots, a_{N} e^{j\phi_N}\}$ represents the diagonal reflection matrix, $a_n \in \{0,1\}$ controls whether the $n$-th element contributes to reflection, and $\phi_n \in [0,2\pi)$ is its phase shift, and the matrix $\mathbf{C}(\mathbf{d}(\boldsymbol{a}))$ models the electromagnetic mutual coupling among RIS elements. Assuming $\mathbf{C}(\mathbf{d}(\boldsymbol{a})) = \mathbf{I}_N + \mathbf{C}_{\text{mutual}}(\mathbf{d}(\boldsymbol{a}))$, where $\mathbf{I}_N$ is the identity matrix and $\mathbf{C}_{\text{mutual}}(\mathbf{d}(\boldsymbol{a}))$ represents the mutual coupling effects determined by the element spacing.

The AP-to-RIS channel vector is defined as 
$\mathbf{h}_{m}(\boldsymbol{a}) = \left[
h_{m,1}(p_1), \ldots, h_{m,N}(p_N)
\right]^{\mathsf{T}}
\in \mathbb{C}^{N \times 1}$, where the positions $\{p_n\}$ are fixed on the grid, while the activation pattern $\boldsymbol{a}$ determines the effective spacing between active elements.
Similarly, $\mathbf{h}_{k}(\boldsymbol{a}) = [h_{1,k}(p_1), \ldots, h_{N,k}(p_N)]^{\mathsf{T}}$ denotes the RIS-to-UE channel vector. The mutual coupling matrix is defined in the Appendix.

Let the effective cascaded channel from all APs to all UEs be defined as  
$\mathbf{G}(\boldsymbol{\Phi}, \boldsymbol{a})
=
\left[ h_{m,k}(\boldsymbol{\Phi}, \boldsymbol{a}) \right]_{m=1,\dots,M}^{k=1,\dots,K}
\in \mathbb{C}^{M \times K}$,
where each entry is given by  
$h_{m,k}(\boldsymbol{\Phi}, \boldsymbol{a})
=
\sqrt{\beta_k}\,
\mathbf{h}_{m}^{\mathsf{H}}(\boldsymbol{a})
\mathbf{C}(\mathbf{d}(\boldsymbol{a}))
\mathbf{R}(\boldsymbol{\Phi}, \boldsymbol{a})
\mathbf{h}_{k}(\boldsymbol{a})$.
The received signal at UE~$k$ is expressed in (\ref{received_signal2}),
\setlength{\dbltextfloatsep}{2pt plus 1pt minus 1pt}
\begin{figure*}[!t]
\centering
\vspace{-1.5em}
\begin{equation}
\label{received_signal2}
\begin{aligned}
y_k
&=
\overbrace{\left(
\sum_{m=1}^{M}
h_{m,k}(\boldsymbol{\Phi}, \boldsymbol{a})
\right)
\sqrt{P_k}\, x_k}^{\text{desired signal}}  
 +
\overbrace{\sum_{\substack{i=1 \\ i \neq k}}^{K}
\left(
\sum_{m=1}^{M}
h_{m,i}(\boldsymbol{\Phi}, \boldsymbol{a})
\right)
\sqrt{P_i}\, x_i}^{\text{inter-user interference}}
+
n_k .
\end{aligned}
\end{equation}
\vspace{-0.5em}
\hrule
\vspace{-0.05em}
\end{figure*}
where $x_i$ denotes the transmitted symbol intended for UE~$i$ with $\mathbb{E}[|x_i|^2]=1$, $P_i$ is the transmit power allocated to UE~$i$, and $n_k \sim \mathcal{CN}(0,\sigma_k^2)$ represents additive white Gaussian noise.
The instantaneous signal-to-interference-plus-noise ratio ($\gamma_k(\boldsymbol{\Phi},\boldsymbol{a})$) at UE~$k$ is $\gamma_k(\boldsymbol{\Phi},\boldsymbol{a})
=
\frac{
P_k
\left|
\sum_{m=1}^{M}
h_{m,k}(\boldsymbol{\Phi}, \boldsymbol{a})
\right|^2
}{
\sum_{\substack{i=1 \\ i \neq k}}^{K} P_i
\left|
\sum_{m=1}^{M}
h_{m,i}(\boldsymbol{\Phi}, \boldsymbol{a})
\right|^2
+
\sigma_k^2
}$, and the achievable rate of UE~$k$ is 
$R_k(\boldsymbol{\Phi}, \boldsymbol{a})
=
\log_2 \left(
1 + \gamma_k(\boldsymbol{\Phi}, \boldsymbol{a})
\right)$.
\subsection{Practical Constraints and Overhead}
A minimum activation constraint $\sum_{n=1}^{N} a_n \geq N_{\min}$ is imposed, where $N_{\min}$ ensures that there are sufficient active elements for beam steering. 
The effective spacing between two successive active elements $n$ and $n+1$ is $d_{\text{eff},n} = (\text{index}_{n+1} - \text{index}_n) \times d_{\min}$, where $\text{index}_n$ is the physical position of the $n$-th active element on the dense grid. This design couples the discrete binary activation decisions with the continuous phase optimization, creating a mixed-integer nonlinear optimization problem.
When incident power hits predominantly one side, this surface alone is powered up, with the other in low-power standby. When UEs are spread on both sides, both surfaces run together to extend coverage and maximize reflection efficiency. 
The overhead of cascaded channel estimation pilots increases linearly with the number of active RIS elements, which is performed once in each coherence block. The practical deployment condition of feasibility is $T_p + T_{\mathrm{opt}} + T_{\mathrm{switch}} < T_c,$
where $T_{\mathrm{switch}}$ is the latency for activation and phase reconfiguration, $T_p$ is the pilot signal length, $T_{\mathrm{opt}}$ is the optimization time, and $T_c$ is the channel coherence. 
The above feasibility condition is based on the assumption of perfect CSI, which might not be the case 
in realistic scenarios. In the case of imperfect CSI, the CSI estimate adds an interference floor in the effective SINR, resulting in a reduction in the achievable rate. For delayed adaptation, the correlation between consecutive coherence blocks is given by $\rho = J_0(2\pi f_D T_c)$, which is $97\%$ effective in the presence of the 
presumed mobility conditions. 

The RIS occupies a finite physical surface; therefore, the total aperture length imposes a constraint on the sum of inter-element spacings as  
\vspace{-0.5em}
\begin{equation}
\sum_{n=1}^{N-1} d_{\text{eff},n} = \sum_{n=1}^{N-1} (\text{index}_{n+1} - \text{index}_n) \, d_{\min} \le D_{\text{total}},
\end{equation}
where $D_{\text{total}}$ denotes the maximum available aperture length. 
\vspace{-0.6em}
\section{Algorithm description}
\vspace{-0,5em}
An adaptive RIS can strengthen useful signal reflections and suppress unwanted ones by jointly tuning its phase shifts and the distance between elements. 
Since effective spacing is determined by activation vector 
$\boldsymbol{a}$, we optimize over $\boldsymbol{a}$ instead of $\boldsymbol{d}$.
CSI acquisition in the double-sided RIS model is a two-step process. In the first phase, pilot signals are sent by APs, and each RIS estimates its dominant propagation direction according to the strength of the received signal.
The activated surface(s) in the second step undergo full CSI estimation using TDD. UEs transmit uplink pilots, which the RIS reflects to APs. The AP estimates the cascaded channel $h_{m,k}(\boldsymbol{\Phi}, \boldsymbol{a})$ by comparing received signals with and without RIS reflection (using ON/OFF keying of the RIS).
The joint phase–spacing optimization problem is represented as 
\vspace{-.5em}
\begin{subequations}
\begin{align}
\max_{\boldsymbol{\Phi}, \boldsymbol{a}} \quad & \min_{k \in \{1,...,K\}} R_k(\boldsymbol{\Phi}, \boldsymbol{a}) \label{eq:obj} \\
\text{s.t.} \quad & \phi_n \in [0, 2\pi), \quad \forall n \in \{1,...,N\}, \label{eq:const_a} \\
& a_n \in \{0, 1\}, \quad \forall n \in \{1,...,N\}, \label{eq:const_b} \\
& \sum_{n=1}^{N} a_n \geq N_{\min}, \label{eq:const_c} \\
& \sum_{n=1}^{N-1} d_{\text{eff},n} \leq D_{\text{total}}. \label{eq:const_d}
\end{align}
\end{subequations}
The objective in (5a) is non-convex due to the min-max structure and cascaded channel dependence $h_{m,k}(\boldsymbol{\Phi},\boldsymbol{a})$, where $\mathbf{R}$ contains both discrete and continuous variables. The feasible set is non-convex due to constraint (5e). NP-hardness follows from reduction to the subset sum problem via constraint (5d). For classical gradient-based optimization, the complexity per iteration is \(\mathcal{O}(N^2K)\) due to the Jacobian computation of \(\nabla_{\boldsymbol{\Phi},\boldsymbol{a}} R_k(\boldsymbol{\Phi},\boldsymbol{a})\). To mitigate the exponential growth of the search space with $N$, QGCN is used that jointly optimizes phase and geometry by using a quantum graph that encodes the RIS. Each element is encoded using dual qubits that represent its binary activation state $a_n \in \{0,1\}$ and continuous phase state $\phi_n \in [0, 2\pi)$.
The RIS is modeled as a graph \(G = (V, E, \mathbf{W})\) with vertices \(V\), edges \(E\), and weight matrix \(\mathbf{W}\). The vertex $v_n$ of the graph is defined by a two-level encoding of a pair of qubits, $|q_n^a\rangle$ and $|q_n^\phi\rangle$. The link weight between elements is $w_{ij} = e^{-\alpha |p_i - p_j|}$, where $(i,j) \in E$, which decreases as the distance between elements increases.
The cost in every iteration is \(\mathcal{O}(S L |E|)\), where \(S = \mathcal{O}(1/\epsilon^2)\) is the number of shots in each measurement to estimate expectation values with precision \(\epsilon\), and \(|E| = \mathcal{O}(N)\) for the sparse graph topology. 
Each parameter requires two circuit evaluations, and the parameter shift rule is used, where $|\boldsymbol{\theta}| = 3L|V_b|$. In subgraph partitioning, $|V_b| \leq$ threshold, and so $|\boldsymbol{\theta}|$ does not depend on $N$. The total complexity of this algorithm is $\mathcal{O}(S L N)$, as opposed to $\mathcal{O}(N^2 K)$ for classical methods with memory $\mathcal{O}(N^2)$.
To overcome the issue of practical scalability due to NISQ hardware limitations, we use graph partitioning to divide the large RIS arrays into subgraphs, each of which is associated with a tractable quantum circuit that may be executed using the available quantum hardware. Assuming that $b$ is the subgraph block index, this approach ensures that the number of qubits required to perform this computation scales as $2|V_b|$ qubits per subgraph rather than the global requirement of $2N$ qubits.
Subgraph solutions are then combined to obtain the global activation pattern $\boldsymbol{a}^*$ and phase profile $\boldsymbol{\Phi}^*$, as described in Algorithm~\ref{alg}. Though this approach to divide and conquer introduces a mild approximation at the boundaries between subgraphs, the ablation results in Table~\ref{ablation} confirm that the full QGCN still maintains considerable benefits even in the noisy regime.
To allow gradients to propagate through the discrete activation choices in quantum circuit training, we first measure the activation qubits during the forward pass to obtain binary states $\{a_n\}$, which define the resulting RIS configuration. Then, during backpropagation, we apply a version of the parameter-shift rule adapted for discrete variables, approximating the gradient with respect to each activation-controlling parameter $\theta_n^a$ as
$\frac{\partial \mathcal{L}}{\partial \theta_n^a} \approx 
\frac{\mathcal{L}\!\big(\theta_n^a + \frac{\pi}{2}\big) - \mathcal{L}\!\big(\theta_n^a - \frac{\pi}{2}\big)}{2}$, where $\mathcal{L}$ is the loss function applied to the quantum circuit output as expressed in (\ref{top}).
To prevent degenerate solutions where all elements are deactivated, we impose a soft constraint in the loss function penalty terms as $\mathcal{L}_{\text{activation}} = \lambda_4 \max\left(0, N_{\min} - \sum_{n=1}^{N_{\text{total}}} a_n\right)^2$,
where $\lambda_4$ is a penalty coefficient and $N_{\min}$ is the minimum number of required active elements to maintain beam steering capability.
Each pair of qubits starts in a uniform superposition $|q_n^{a}\rangle = |1\rangle, \quad
|q_n^{\phi}\rangle = |+\rangle = \frac{1}{\sqrt{2}}(|0\rangle + |1\rangle)$.
We link phase and activation using entanglement as
\vspace{-6pt}
\begin{equation}
|\psi_0\rangle = \prod_{n=1}^{|V_b|} \text{CNOT}(q_n^a, q_n^{\phi})
\bigotimes_{n=1}^{|V_b|}
\left(|1\rangle_n^{a} \otimes |+\rangle_n^{\phi}\right).
\end{equation}
\vspace{-6pt}
After encoding, the graph state is expressed as
\begin{equation}
|\psi_0\rangle =
\prod_{(i,j)\in E}\text{CNOT}_{ij}^{\phi}
\prod_{n=1}^{|V_b|}\text{CNOT}(q_n^a,q_n^{\phi})
\bigotimes_{n=1}^{|V_b|}
\left(|1\rangle_n^{a}\otimes|+\rangle_n^{\phi}\right).
\end{equation}
Neighboring elements interact by a controlled-phase operation weighted by distance as $U_{GC}^{(l)}(\boldsymbol{\theta}^{(l)}, \mathbf{W}) =
\prod_{(i,j)\in E_b}\text{CPhase}_{ij}\big(w_{ij}\theta_{ij}^{(l)}\big)$.
The controlled-phase gate acts only when both qubits corresponding to elements 
$i$ and $j$ (edge endpoints) are in the excited state.
After these interactions, the overall system moves from the previous state $|\psi^{(l-1)}\rangle$ to a new configuration as 
$|\psi^{(l)}\rangle=U_{GC}^{(l)}|\psi^{(l-1)}\rangle$, which now carries phase correlations that depend on spacing.
To enable densely packed regions to adjust positions cooperatively rather than independently, each element updates its own spacing using a small rotation influenced by its neighbors' average phase, which is expressed as $U_{S}^{(l)}=\prod_{i}R_y^{(i,a)}\!\Big(\beta_i^{(l)}\!\sum_{j\in\mathcal{N}(i)}\!\langle Z_j^{\phi}\rangle\Big)$.
Both phase and spacing qubits undergo parameterized rotations as $R_{yz}(\alpha_i,\gamma_i)=R_z(\gamma_i)R_y(\alpha_i).$ Combining these rotations for all elements gives
$U_{A}^{(l)}=\bigotimes_{i}\!\left[R_{yz}(\alpha_i^{(l)},\gamma_i^{(l)})\otimes R_y(\beta_i^{(l)})\right]$.  
The updated system state is $|\psi^{(l)}\rangle=U_{A}^{(l)}U_{S}^{(l)}U_{GC}^{(l)}|\psi^{(l-1)}\rangle$.
Stacking several of these operations produces a deep adaptive circuit as
$|\psi_L\rangle=\prod_{l=1}^{L}\!\left[U_{A}^{(l)}U_{S}^{(l)}U_{GC}^{(l)}\right]\!|\psi_0\rangle$,
where coupling weights are refreshed after each layer to reflect the latest estimated positions defined as $w_{ij}^{(l+1)}=e^{-\alpha|\hat{p}_i^{(l)}-\hat{p}_j^{(l)}|}$, and
$\hat{p}_n^{(l)}=\sum_{k=1}^{n-1}\langle a_k^{(l)}\rangle d_{\min}$.
At the end of the circuit, qubits are measured to obtain the physical quantities of interest as $a_n = \text{measurement of } q_n^{a}$, and
$\phi_n=2\pi\langle Z_n^{\phi}\rangle$,
which provides the final activation states and phase shifts.
\begin{figure*}[t]
\vspace{-4.5pt}
\begin{equation}
\label{top}
\begin{aligned}
\mathcal{L}=&-\sum_k w_kR_k(\boldsymbol{\Phi},\boldsymbol{a})
+\lambda_1\|\boldsymbol{\Phi}\|_F^2
+\lambda_2\mathbb{I}\!\left(\sum_{n=1}^{N-1} d_{\text{eff},n}(\boldsymbol{a})>D_{\text{total}}\right) 
&+ \lambda_3 \max\left(0, N_{\min} - \sum_{n=1}^{N} a_n\right)^2
\end{aligned}
\end{equation}
\vspace{-2pt}
\hrule
\end{figure*}
\vspace{-2pt}
The summary of the algorithm is explained at Algorithm \ref{alg}.

\begin{algorithm}[!htb]
\caption{QGCN Algorithm}
\label{alg}
\scriptsize
\begin{algorithmic}[1]
\Require $G=(V,E)$, $\{h_{m,k}\}$, $d_{\min}$, $D_{\text{total}}$, $N_{\min}$, L, T, $\eta$
\Ensure $\boldsymbol{\Phi}^*$, $\boldsymbol{a}^*$, $\mathbf{d}_{\text{eff}}^*$
\State Partition $G$ into subgraphs with $|V_b| \leq 40$
\For{each $G_b$}
 \State $\boldsymbol{\theta}_b = \{\alpha_n^{(\ell)}, \beta_n^{(\ell)}, \gamma_n^{(\ell)}\}$; \quad $|q_n^a\rangle \gets |1\rangle$, $|q_n^\phi\rangle \gets |+\rangle$, $\forall n \in V_b$
 \For{$t = 1$ to $T$}
 \State $|\psi\rangle \gets \text{CNOT}(q_n^a, q_n^\phi)\,|\psi\rangle$, $\forall n \in V_b$
 \For{$\ell = 1$ to $L$}
 \State $|\psi\rangle \gets \text{CPhase}(w_{ij})\,|\psi\rangle$, $\forall (i,j) \in E_b$
 \For{$n \in V_b$}
 \State $\mu_n \gets \sum_{j \in \mathcal{N}(n)} \langle Z_j^\phi \rangle$
 \State $|\psi\rangle \gets R_z(\gamma_n^{(\ell)}) R_y(\alpha_n^{(\ell)} + \beta_n^{(\ell)} \mu_n)\,|\psi\rangle$
 \EndFor
 \EndFor
 \State $a_n \gets \text{measure}(q_n^a)$, $\forall n \in V_b$
 \State Evaluate $\{p_n\}$ and $w_{ij} \gets \exp(-\kappa |p_i - p_j|)$, $\forall (i,j) \in E_b$
 \State Compute gradients $\nabla_{\boldsymbol{\theta}_b}$ and update $\boldsymbol{\theta}_b$
 \EndFor
 \State Save $(\boldsymbol{a}_b^*, \boldsymbol{\Phi}_b^*)$
\EndFor
\State Connect all subgraphs to get $(\boldsymbol{a}^*, \boldsymbol{\Phi}^*)$ and $\mathbf{d}_{\text{eff}}^*$
\State \Return $\boldsymbol{\Phi}^*, \boldsymbol{a}^*, \mathbf{d}_{\text{eff}}^*$
\end{algorithmic}
\end{algorithm}
\vspace{-0.3em}
\section{Simulation Results}
\vspace{-0.3em}
We evaluated the performance of the proposed quantum optimization algorithm using both simulation (PennyLane) and quantum hardware on IBM Quantum's ibm\_kyiv backend (127-qubit superconducting quantum processor) with parameters shown in Table \ref{params}. 
This $\mathcal{O}(LN)$ gate complexity with depth $\mathcal{O}(L|E|)$ enables NISQ implementation.
\begin{table}[htbp]
\vspace{-0.7em}
\centering
\scriptsize
\caption{Simulation Parameters}
\label{params}
\setlength{\tabcolsep}{1.5pt}  
\begin{tabular}{@{}llllllll@{}}
\hline
\textbf{Param.} & \textbf{Val.} & \textbf{Param.} & \textbf{Val.} & \textbf{Param.} & \textbf{Val.} & \textbf{Param.} & \textbf{Val.} \\
\hline
Carrier freq. & 28 GHz & Bandwidth & 100 MHz & Rician $\kappa$ & 10 dB & User vel. & 3 m/s \\
Doppler $f_D$ & 280 Hz & Coherence & 1.5 ms & Pilot time & 0.1 ms & Opt. time & 18.3 ms \\
Symbol dur. & 10 $\mu$s & Noise pwr. & -94 dBm & RIS elem. & 20--100 & UEs ($K$) & 4--20 \\
APs ($M$) & 2--4 & Path loss & 2.2,3.8 & Min. space & $\lambda$/2 & Layers $L$ & 4 \\
Tx power & 1 W & Learn. rate & 0.01 & SNR & 10 dB & MC runs & 50 \\
Backend & \multicolumn{3}{l}{IBM Kyiv} & Qubits & 127 & & \\
\hline
\end{tabular}
\end{table}
\vspace{-6pt}
\begin{figure*}[!ht]
\centering
\includegraphics[height = 1.5in, width=6in]{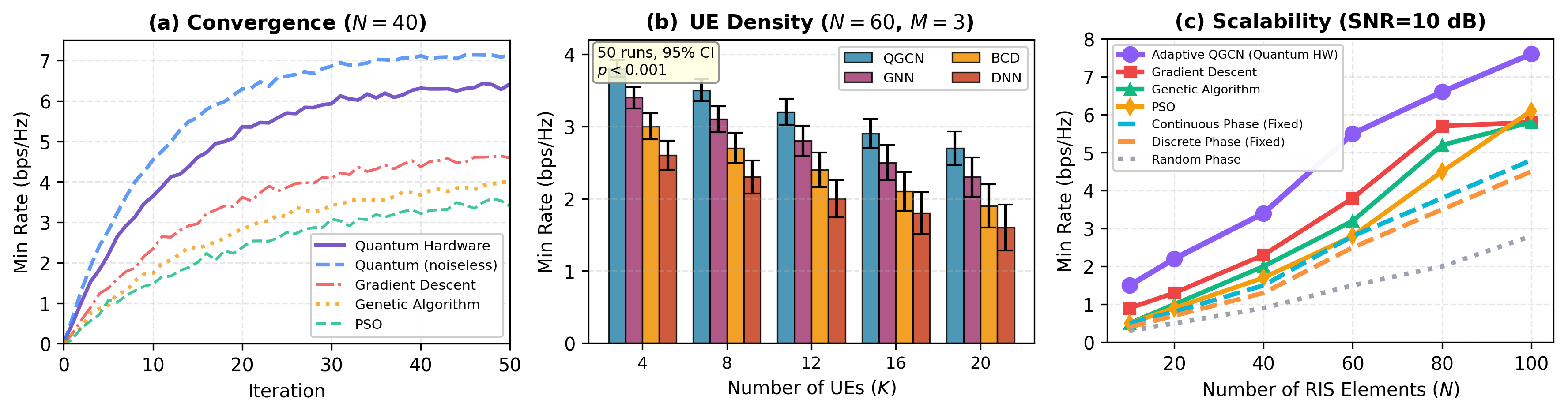}
\caption{(a) Quantum hardware achieves 89\% of noiseless simulation with faster convergence. (b) Scalability at 95\% CI, simulated. (c) Minimum user rate, near-linear growth, simulated.}
\label{quantum_validation}
\end{figure*} \vspace{0.5em}
Fig.~\ref{quantum_validation} presents the validation results, combining quantum hardware execution (Fig.~\ref{quantum_validation}(a)) and simulation with 50 Monte Carlo runs (Fig.~\ref{quantum_validation}(b-c)). We compare against continuous-phase, discrete-phase, random-phase, and fixed-spacing, as well as 
classical baselines, including gradient descent (GD), genetic algorithm (GA), particle swarm optimization (PSO), block coordinate descent (BCD), GNN, and 
DNN. As far as scalability is concerned, the subgraph partitioning strategy enables QGCN to process large arrays within the current hardware constraints. Fig.~\ref{quantum_validation}(a) shows 
that the quantum hardware implementation 
converges to $89\%$ of noiseless simulation 
performance within 30 iterations, 
outpacing classical baselines. Fig. \ref{quantum_validation}(b) provides a quantification of the incremental gain of the proposed QGCN algorithm over classical learning techniques. Classical GNN uses the same graph structure as QGCN, where the weights of the edges are the same. It uses a continuous relaxation of the binary activation variables. In Fig.~\ref{quantum_validation}(c), the minimum user rate increases almost linearly with $N$. 
The performance gain of QGCN over the classical GNN is +0.38 bps/Hz for N = 60. This gap is further widened for larger arrays as the discrete activation space grows exponentially. The gain is not only a result of the architecture; it is larger where continuous relaxation leads to a larger error.
To evaluate the effect of every architectural component, we perform an ablation study using Qiskit Aer simulator with noise models set to IBM \texttt{ibm\_marrakesh} (Heron~r2, 156 qubits), as depicted in Table~\ref{ablation}. The noise model includes single-qubit depolarizing error (0.1\%), two-qubit depolarizing error (0.7\%) applied to CX and CP gates, and symmetric readout bit-flip error (1.5\%). The three settings are: (A) Full QGCN, (B) QGCN without virtual spacing (all elements enabled), and (C) QGCN without double-sided RIS (half of the elements disabled).
Also, Table~\ref{complexity} summarizes the per-iteration complexity and memory requirements of QGCN against classical baselines.
\begin{table}[htbp]
\centering
\caption{Ablation results confirming virtual spacing benefits by increasing spatial diversity. N = 4, M = 2, K = 3, L = 2, SNR = 10 dB, 2048 shots, 5 Monte Carlo runs, 95\% CI.}
\label{ablation}
\begin{tabular}{@{}lccc@{}}
\hline
\textbf{Configuration} & \textbf{Min.\ Rate (bps/Hz)} & \textbf{Degrad.} & \textbf{Conv.\ Epoch} \\
\hline
Full QGCN (baseline)   & $0.139 \pm 0.122$ & ---          & $13 \pm 9$ \\
w/o Virtual Spacing    & $0.104 \pm 0.084$ & $-25.1\%$    & $13 \pm 9$ \\
w/o Double-Sided RIS   & $0.089 \pm 0.009$ & $-35.9\%$    & $6  \pm 7$ \\
\hline
\end{tabular}
\end{table}
\vspace{-12pt}
\begin{table}[htbp]
\centering
\scriptsize
\caption{Complexity comp. : $S=2048$, $L=4$, $\epsilon=0.01)$}.
\label{complexity}
\setlength{\tabcolsep}{2pt}
\begin{tabular}{@{}lccc@{}}
\hline
\textbf{Method} & 
\textbf{Per-iter.} & 
\textbf{Memory} & 
\textbf{Noise} \\
 & 
\textbf{Complexity} & 
\textbf{Complexity} & 
\textbf{Overhead} \\
\hline
QGCN (proposed) & 
$\mathcal{O}(S{\cdot}L|E|)$ & 
Implicit$^\dagger$ & 
$2|\boldsymbol{\theta}|$ circuits \\
Classical GNN & 
$\mathcal{O}(N^2K)$ & 
$\mathcal{O}(N^2)$ & 
None \\
BCD & 
$\mathcal{O}(N^3)$ & 
$\mathcal{O}(N^2)$ & 
None \\
DNN & 
$\mathcal{O}(N^2)$ & 
$\mathcal{O}(N^2)$ & 
None \\
\hline
\multicolumn{4}{@{}l}{%
$^\dagger$Pairwise interactions encoded implicitly in quantum state;} \\
\multicolumn{4}{@{}l}{%
no explicit $\mathcal{O}(N^2)$ storage required.} \\
\hline
\end{tabular}
\end{table}
\vspace{0.5em}
\section{Conclusion and Future Work}
\vspace{-0.4em}
As a critical technology to sculpting wireless propagation in next-generation networks, RISs have received growing attention.
In this paper, we tackled the scalability of double-sided RIS from two perspectives: algorithm and hardware design by introducing a switch-controlled virtual spacing architecture. 
The proposed algorithm achieves higher minimum UE rates and faster convergence. Beyond simulation, we validated the algorithm on an IBM Quantum processor, where the quantum implementation achieves 89\% of noiseless simulation performance.
Next-generation processors with more qubits and improved coherence are expected to support larger deployments. 
\vspace{-18pt}
\section*{Appendix}
\vspace{-0.3em}
For the coupling matrix $\mathbf{C}(\mathbf{d}(\boldsymbol{a}))$, reducing the inter-element spacing strengthens mutual coupling, thereby altering the effective reflection coefficients and distorting the intended beam pattern. The mutual coupling component is defined as  
\textbf{$[\mathbf{C}_{\text{mutual}}(\mathbf{d}(\boldsymbol{a}))]_{ij} =
\begin{cases}
0, & i = j, \\
\frac{\sin(k_0 |p_i - p_j|)}{k_0 |p_i - p_j|} e^{-j k_0 |p_i - p_j|}, & i \neq j,
\end{cases}$}
where $p_i$ and $p_j$ are the element positions and $k_0$ is the wave number.
As spacing decreases, the increased mutual coupling progressively distorts the effective reflection response and beamforming pattern.
\vspace{-6pt}
\bibliography{IEEEabrv,cite}
\vspace{-3pt}

\end{document}